
\font\gross=cmbx10  scaled\magstep2
\font\mittel=cmbx10 scaled\magstep1
\def\gsim{\mathrel{\raise.3ex\hbox{$>$\kern- .75em
                      \lower1ex\hbox{$\sim$}}}}
\def\lsim{\mathrel{\raise.3ex\hbox{$<$\kern-.75em
                      \lower1ex\hbox{$\sim$}}}}

\def\square{\kern 1pt
\vbox{\hrule height 0.6pt\hbox{\vrule width 0.6pt \hskip 3pt
\vbox{\vskip 6pt}\hskip 3pt\vrule width 0.6pt}
\hrule height 0.6pt}\kern 1pt }

\def\sla{\raise.15ex\hbox{$/$}\kern-.72em}

\def\singlespace{\baselineskip=\normalbaselineskip}

\parskip=\medskipamount
\overfullrule=0pt
\raggedbottom
\def\normalparindent{24pt}
\newif\ifdraft \draftfalse

\nopagenumbers
\footline={\ifnum\pageno=1 {\ifdraft
{\hfil\rm Draft \number\day -\number\month -\number\year}
\else{\hfil}\fi}
\else{\hfil\rm\folio\hfil}\fi}
\def\endpage{\vfill\eject}
\def\beginlinemode{\endmode\begingroup\parskip=0pt
\obeylines\def\\{\par}\def\endmode{\par\endgroup}}
\def\beginparmode{\endmode\begingroup \def\endmode{\par\endgroup}}
\let\endmode=\par
\def\raggedcenter{
                  \leftskip=2em plus 6em \rightskip=\leftskip
                  \parindent=0pt \parfillskip=0pt \spaceskip=.3333em
                  \xspaceskip=.5em\pretolerance=9999 \tolerance=9999
                  \hyphenpenalty=9999 \exhyphenpenalty=9999 }
\def\\{\cr}
\let\rawfootnote=\footnote\def\footnote#1#2{{\parindent=0pt\parskip=0pt
        \rawfootnote{#1}{#2\hfill\vrule height 0pt depth 6pt width 0pt}}}
\def\title{\null\vskip 3pt plus 0.2fill\beginlinemode\raggedcenter\gross}
\def\author{\vskip 3pt plus 0.2fill \beginlinemode\raggedcenter}
\def\affil{\vskip 3pt plus 0.1fill\beginlinemode\raggedcenter\it}
\def\abstract{\vskip 3pt plus 0.3fill \beginparmode{\noindent
{\mittel Abstract}:~}  }
\def\endtitlepage{\endpage\body}
\def\body{\beginparmode\parindent=\normalparindent}
\def\head#1{\par\goodbreak{\immediate\write16{#1}
      \vskip 0.4cm{\noindent\gross #1}\par}\nobreak\nobreak\nobreak\nobreak}

\def\finalcite{\citeall\ref\citeall\Ref}
\newif\ifannpstyle
\newif\ifprdstyle
\newif\ifplbstyle
\newif\ifwsstyle

\gdef\refto#1{\ifprdstyle  $^{\[#1] }$ \else
              \ifwsstyle$^{\[#1]}$  \else
              \ifannpstyle $~[\[#1] ]$ \else
              \ifplbstyle  $~[\[#1] ]$ \else
                                         $^{[\[#1] ]}$\fi\fi\fi\fi}
\gdef\refis#1{\ifprdstyle \item{~$^{#1}$}\else
              \ifwsstyle \item{#1.} \else
              \ifplbstyle\item{~[#1]} \else
              \ifannpstyle \item{#1.} \else
                              \item{#1.\ }\fi\fi\fi\fi }
\gdef\journal#1,#2,#3,#4.{
           \ifprdstyle {#1~}{\bf #2}, #3 (#4).\else
           \ifwsstyle {\it #1~}{\bf #2~} (#4) #3.\else
           \ifplbstyle {#1~}{#2~} (#4) #3.\else
           \ifannpstyle {\sl #1~}{\bf #2~} (#4), #3.\else
                       {\sl #1~}{\bf #2}, #3 (#4)\fi\fi\fi\fi}

\def\ref#1{Ref.~#1}
\def\Ref#1{Ref.~#1}
\def\cite#1{{#1}}\def\[#1]{\cite{#1}}

\def\eq#1{Eq.~\(#1)}\def\eqs#1{Eqs.~\(#1)}

\def\(#1){(\call{#1})}
\def\call#1{{#1}}\def\taghead#1{{#1}}

\def\references{\head{References}\beginparmode\frenchspacing\parskip=0pt}
\def\endreferences{\body}
\def\endit{\endmode\vfill\supereject}\let\endpaper=\endit


\def\konstanz{Universit\"at Konstanz, Fakult\"at f\"ur Physik,\\
Postfach 5560, D-7750 Konstanz, Germany.\\E-mail: phlousto@dknkurz1}

\def\infn{On leave from the Dipartimento di Fisica, Universit\`a di Perugia,
Italy. \\ \noindent E-mail: phsirio@dknkurz1}

\def\iafedir{Permanent Adress: Instituto de Astronom\'\i a y F\'\i sica del
Espacio, Casilla de Correo 67 -\\ Sucursal 28, 1428 Buenos Aires, Argentina.
 E-mail: lousto@iafe.edu.ar}

\def\mangos{This work was partially supported
by the Directorate General for
Science, Research and Development of
the Commission of the European Communities
and by the Alexander von Humboldt Foundation.}

\catcode`@=11
\newcount\r@fcount \r@fcount=0\newcount\r@fcurr
\immediate\newwrite\reffile\newif\ifr@ffile\r@ffilefalse
\def\w@rnwrite#1{\ifr@ffile\immediate\write\reffile{#1}\fi\message{#1}}
\def\writer@f#1>>{}
\def\referencefile{\r@ffiletrue\immediate\openout\reffile=\jobname.ref%
  \def\writer@f##1>>{\ifr@ffile\immediate\write\reffile%
    {\noexpand\refis{##1} = \csname r@fnum##1\endcsname = %
     \expandafter\expandafter\expandafter\strip@t\expandafter%
     \meaning\csname r@ftext\csname r@fnum##1\endcsname\endcsname}\fi}%
  \def\strip@t##1>>{}}

\def\citeall#1{\xdef#1##1{#1{\noexpand\cite{##1}}}}
\def\cite#1{\each@rg\citer@nge{#1}}
\def\each@rg#1#2{{\let\thecsname=#1\expandafter\first@rg#2,\end,}}
\def\first@rg#1,{\thecsname{#1}\apply@rg}
\def\apply@rg#1,{\ifx\end#1\let\next=\relax%
\else,\thecsname{#1}\let\next=\apply@rg\fi\next}%
\def\citer@nge#1{\citedor@nge#1-\end-}
\def\citer@ngeat#1\end-{#1}
\def\citedor@nge#1-#2-{\ifx\end#2\r@featspace#1
  \else\citel@@p{#1}{#2}\citer@ngeat\fi}
\def\citel@@p#1#2{\ifnum#1>#2{\errmessage{Reference range #1-#2\space is bad.}
\errhelp{If you cite a series of references by the notation M-N, then M and
    N must be integers, and N must be greater than or equal to M.}}\else%
{\count0=#1\count1=#2\advance\count1
by1\relax\expandafter\r@fcite\the\count0,%
  \loop\advance\count0 by1\relax
    \ifnum\count0<\count1,\expandafter\r@fcite\the\count0,%
  \repeat}\fi}
\def\r@featspace#1#2 {\r@fcite#1#2,}    \def\r@fcite#1,{\ifuncit@d{#1}
    \expandafter\gdef\csname r@ftext\number\r@fcount\endcsname%
{\message{Reference #1 to be supplied.}\writer@f#1>>#1 to be supplied.\par
     }\fi\csname r@fnum#1\endcsname}
\def\ifuncit@d#1{\expandafter\ifx\csname r@fnum#1\endcsname\relax%
\global\advance\r@fcount by1%
\expandafter\xdef\csname r@fnum#1\endcsname{\number\r@fcount}}
\let\r@fis=\refis   \def\refis#1#2#3\par{\ifuncit@d{#1}%
\w@rnwrite{Reference #1=\number\r@fcount\space is not cited up to now.}\fi%
  \expandafter\gdef\csname r@ftext\csname r@fnum#1\endcsname\endcsname%
  {\writer@f#1>>#2#3\par}}
\def\r@ferr{\endreferences\errmessage{I was expecting to see
\noexpand\endreferences before now;  I have inserted it here.}}
\let\r@ferences=\references
\def\references{\r@ferences\def\endmode{\r@ferr\par\endgroup}}
\let\endr@ferences=\endreferences
\def\endreferences{\r@fcurr=0{\loop\ifnum\r@fcurr<\r@fcount
\advance\r@fcurr by 1\relax\expandafter\r@fis\expandafter{\number\r@fcurr}%
    \csname r@ftext\number\r@fcurr\endcsname%
  \repeat}\gdef\r@ferr{}\endr@ferences}
\let\r@fend=\endpaper\gdef\endpaper{\ifr@ffile
\immediate\write16{Cross References written on []\jobname.REF.}\fi\r@fend}
\catcode`@=12
\finalcite
\catcode`@=11
\newcount\tagnumber\tagnumber=0
\immediate\newwrite\eqnfile\newif\if@qnfile\@qnfilefalse
\def\write@qn#1{}\def\writenew@qn#1{}
\def\w@rnwrite#1{\write@qn{#1}\message{#1}}
\def\@rrwrite#1{\write@qn{#1}\errmessage{#1}}
\def\taghead#1{\gdef\t@ghead{#1}\global\tagnumber=0}
\def\t@ghead{}\expandafter\def\csname @qnnum-3\endcsname
  {{\t@ghead\advance\tagnumber by -3\relax\number\tagnumber}}
\expandafter\def\csname @qnnum-2\endcsname
  {{\t@ghead\advance\tagnumber by -2\relax\number\tagnumber}}
\expandafter\def\csname @qnnum-1\endcsname
  {{\t@ghead\advance\tagnumber by -1\relax\number\tagnumber}}
\expandafter\def\csname @qnnum0\endcsname
  {\t@ghead\number\tagnumber}
\expandafter\def\csname @qnnum+1\endcsname
  {{\t@ghead\advance\tagnumber by 1\relax\number\tagnumber}}
\expandafter\def\csname @qnnum+2\endcsname
  {{\t@ghead\advance\tagnumber by 2\relax\number\tagnumber}}
\expandafter\def\csname @qnnum+3\endcsname
  {{\t@ghead\advance\tagnumber by 3\relax\number\tagnumber}}
\def\equationfile{\@qnfiletrue\immediate\openout\eqnfile=\jobname.eqn%
  \def\write@qn##1{\if@qnfile\immediate\write\eqnfile{##1}\fi}
  \def\writenew@qn##1{\if@qnfile\immediate\write\eqnfile
    {\noexpand\tag{##1} = (\t@ghead\number\tagnumber)}\fi}}
\def\callall#1{\xdef#1##1{#1{\noexpand\call{##1}}}}
\def\call#1{\each@rg\callr@nge{#1}}
\def\each@rg#1#2{{\let\thecsname=#1\expandafter\first@rg#2,\end,}}
\def\first@rg#1,{\thecsname{#1}\apply@rg}
\def\apply@rg#1,{\ifx\end#1\let\next=\relax%
\else,\thecsname{#1}\let\next=\apply@rg\fi\next}
\def\callr@nge#1{\calldor@nge#1-\end-}\def\callr@ngeat#1\end-{#1}
\def\calldor@nge#1-#2-{\ifx\end#2\@qneatspace#1 %
  \else\calll@@p{#1}{#2}\callr@ngeat\fi}
\def\calll@@p#1#2{\ifnum#1>#2{\@rrwrite{Equation range #1-#2\space is bad.}
\errhelp{If you call a series of equations by the notation M-N, then M and
N must be integers, and N must be greater than or equal to M.}}\else%
{\count0=#1\count1=#2\advance\count1 by1
\relax\expandafter\@qncall\the\count0,%
  \loop\advance\count0 by1\relax%
    \ifnum\count0<\count1,\expandafter\@qncall\the\count0,  \repeat}\fi}
\def\@qneatspace#1#2 {\@qncall#1#2,}
\def\@qncall#1,{\ifunc@lled{#1}{\def\next{#1}\ifx\next\empty\else
\w@rnwrite{Equation number \noexpand\(>>#1<<) has not been defined yet.}
 >>#1<<\fi}\else\csname @qnnum#1\endcsname\fi}
\let\eqnono=\eqno\def\eqno(#1){\tag#1}\def\tag#1$${\eqnono(\displayt@g#1 )$$}
\def\aligntag#1\endaligntag  $${\gdef\tag##1\\{&(##1 )\cr}\eqalignno{#1\\}$$
  \gdef\tag##1$${\eqnono(\displayt@g##1 )$$}}
\def\eqalignno#1{\displ@y \tabskip\centering
  \halign to\displaywidth{\hfil$\displaystyle{##}$\tabskip\z@skip
    &$\displaystyle{{}##}$\hfil\tabskip\centering
    &\llap{$\displayt@gpar##$}\tabskip\z@skip\crcr
    #1\crcr}}
\def\displayt@gpar(#1){(\displayt@g#1 )}
\def\displayt@g#1 {\rm\ifunc@lled{#1}\global\advance\tagnumber by1
        {\def\next{#1}\ifx\next\empty\else\expandafter
        \xdef\csname @qnnum#1\endcsname{\t@ghead\number\tagnumber}\fi}%
  \writenew@qn{#1}\t@ghead\number\tagnumber\else
        {\edef\next{\t@ghead\number\tagnumber}%
        \expandafter\ifx\csname @qnnum#1\endcsname\next\else
        \w@rnwrite{Equation \noexpand\tag{#1} is a duplicate number.}\fi}%
  \csname @qnnum#1\endcsname\fi}
\def\eqnoa(#1){\global\advance\tagnumber by1\multitag{#1}{a}}
\def\eqnob(#1){\multitag{#1}{b}}
\def\eqnoc(#1){\multitag{#1}{c}}
\def\eqnod(#1){\multitag{#1}{d}}
\def\multitag#1#2$${\eqnono(\multidisplayt@g{#1}{#2} )$$}
\def\multidisplayt@g#1#2 {\rm\ifunc@lled{#1}
        {\def\next{#1}\ifx\next\empty\else\expandafter
        \xdef\csname @qnnum#1\endcsname{\t@ghead\number\tagnumber b}\fi}%
  \writenew@qn{#1}\t@ghead\number\tagnumber #2\else
        {\edef\next{\t@ghead\number\tagnumber #2}%
        \expandafter\ifx\csname @qnnum#1\endcsname\next\else
\w@rnwrite{Equation \noexpand\multitag{#1}{#2} is a duplicate number.}\fi}%
  \csname @qnnum#1\endcsname\fi}
\def\ifunc@lled#1{\expandafter\ifx\csname @qnnum#1\endcsname\relax}
\let\@qnend=\end\gdef\end{\if@qnfile
\immediate\write16{Equation numbers written on []\jobname.EQN.}\fi\@qnend}

\singlespace
\magnification=1200
\vskip 24pt

 \title

  Are  Black Holes in Brans-Dicke Theory

  precisely the same as in General Relativity?

\author

M.Campanelli\footnote{$^{*}$}{\infn} and
C. O. Lousto\footnote{$^{**}$}{\iafedir}

\affil
\konstanz

\abstract

We study a three-parameters family of solutions of the Brans-Dicke field
equations. They are static and spherically symmetric. We find the range
of  parameters for which this solution represents a black hole different
from the Schwarzschild one. We find a subfamily of solutions which agrees
with experiments and observations in the solar system. We discuss some
astrophysical applications and the consequences on the "no hair"
theorems for black holes.

\bigskip
\bigskip
\bigskip

PACS = 04.50.+h

\endtitlepage
\singlespace

\head{1. Introduction}

Lately there have been some renewed interest in the Brans-Dicke
theory of gravitation. On one hand, it has been applied to
cosmological models of the universe during the inflationary era
to make more natural bubble percolation \refto{1}.
Also, it was found that in the low-energy regime, the theory of
fundamental strings can be reduced to an effective Brans-Dicke
one\refto{2}.
The subject of gravitational collapse, however,  has not yet been
thoroughly studied. One of the outstanding results on this field is
the Hawking theorem\refto{3},  that states that the Schwarzschild
metric is the only spherically symmetric solution of vacuum Brans-
Dicke field equations.
The proof of this theorem goes through the fact that the Brans-Dicke
scalar field $\phi$  must be constant  outside the black hole and the
use of the weak energy condition. In this
paper we study a three-parameters family of solutions of Brans-Dicke
equations which is  static and spherically symmetric. We study under
which range of the parameters we can have non-singular (at the horizon)
black hole solutions. We are able to obtain explicit examples where
the metric represents a black hole solution different from the
Schwarzschild one:
$$
ds^2=-A(r)^{1-n}dt^2+A(r)^{n-1}dr^2+r^2A(r)^nd\Omega^2 ~~;
{}~~A(r)=1-2{r_0\over r}~~; ~~n\leq -1~~.      \eqno(58)
$$
where $r_0$ is an arbitrary constant and $n$ represents a scalar hair.

Classical scalar hairs in General Relativity Black Hole solutions have
already been found for several coupling. These include the case of an
axion with an $R\tilde R$ coupling \refto{CDKO}, and a dilaton
\refto{GHS} and an axionlike scalar field \refto{LW} coupled to Einstein
Maxwell theory.
A conformally coupled scalar field can have a static \refto{B}, but
unstable \refto{BK} solution.

The Brans-Dicke theory \refto{4} incorporates the Mach principle, which
states that the phenomenon of inertia  must arise from accelerations with
respect to the general mass distribution of the universe. This theory is
self-consistent, complete and for $|\omega|\geq 500$ in accord with solar
system observations and experiments \refto{5}.
It is, in some sense, the  simplest extension of General Relativity.
It introduces an additional long-range scalar field $\phi$
besides the metric tensor of the spacetime $g_{\mu\nu}$ from which are
constructed the covariant derivative and the curvature tensors, in the
usual manner. $\omega$ is the Dicke dimensionless coupling constant.

The theory is metric, i. e. the weak equivalence principle is satisfied.
The matter couples minimally to the metric and not directly to $\phi$.
The scalar field does not exert any direct influence on matter, its only
role is that of participating in the field equations that determine the
geometry of the spacetime.

The action for the Brans-Dicke theory is:
$$
S=\intop\nolimits{dx^4\sqrt{-g}[\phi R - \omega(\phi_{,\alpha}
\phi^{,\alpha})/\phi + 16\pi L_{matter}]} ~~.                  \eqno(1)
$$
The variational principle gives the field equations:
$$
G_{\alpha\beta}={8\pi\over\phi}T_{\alpha\beta}+{\omega\over\phi^2}
\left(\phi_{,\alpha}\phi_{,\beta} -{1\over 2}g_{\alpha\beta}
\phi_{,\mu}\phi^{,\mu}\right)+ {1\over\phi}(\phi_{;\alpha\beta}-
g_{\alpha\beta}\square\phi)~~.                                 \eqno(2)
$$
The matter stress-energy tensor and $\phi$ together generate the metric.
The field equation for $\phi$ is:
$$
\phi_{; \alpha}^{ ~~ \alpha}=\square\phi={8\pi\over 3+2\omega}T ~~.
                                                               \eqno(3)
$$
In the next section we study a solution\refto{4} of the vacuum field
\eqs{2}-\(3). This is a three-parameters static spherically symmetric
metric. We study the asymptotic behavior, the occurrence of singularities
and event horizons. In the third section we study the special cases for
which this metric can be of astrophysical relevance. We compute the
geodesics equations, post Newtonian parameters, energy, period and
redshift of the last stable circular orbit, dispersion cross sections,
Kruskal transformations and Hawking temperature. We end the paper with the
discussion of the obtained results,  in particular, the relevance of the
non-Schwarzschild-like black holes found in the Brans-Dicke theory.

\bigskip

\head{2. Static Spherically Symmetric Vacuum Solutions}

The Brans-Dicke vacuum field equations can be written as:
$$
R_{\alpha\beta}= {\omega \over \phi^2}\phi_{,\alpha}\phi_{,\beta}+
{\phi_{;\alpha\beta}\over\phi}~~,                              \eqno(4)
$$
$$
\square\phi=0 ~~.                                              \eqno(4')
$$
It is easy to show that a power   generalization of the Schwarzschild
metric is a solution of this equations \refto{4, 6}:
$$
ds^2=- A(r)^{m+1} dt^2 + A(r)^{n-1} dr^2 + r^2 A(r)^n d\Omega^2  ~~;
  \eqno(6)
$$
$$
d\Omega^2=d\vartheta^2 + \sin^2\vartheta\varphi^2 ~~;
$$
$$
A(r)=1-2{r_0\over r}~~,
$$
and the scalar field:
$$
\phi(r)=\phi_0 A(r)^{-{m+n\over 2}} ~~;                \eqno(7)
$$
where $m$, $n$, $\phi_0$ and $r_0$ are arbitrary constants.
The coupling constant is found from:
$$
\omega=-2{(m^2+n^2+nm+m-n)\over (m+n)^2} ~~.        \eqno(7')
$$
Either from \(6) or \(7) we can compute the components of the Ricci
tensor:
$$
R_{00}=(m+1)(m+n){r_0^2\over r^4} A(r)^{m-n} ~~,            \eqno(8)
$$

$$
R_{11}=(-m^2+nm+3n-m){r_0^2\over r^4}A(r)^{-2}+2(m+n){r_0\over r^3}
A(r)^{-1} ~~, \eqno(9)
$$

$$
R_{22}=-n(m+n)\left({r_0^2\over r^2}\right)A(r)^{-1}-(m+n){r_0\over r} ~~,
     \eqno(10)
$$
$$
R_{33}=\sin^2\vartheta R_{22} ~~,                             \eqno(11)
$$
$$
R_{\mu\nu}=0  ~~~~~~~~~(\mu\neq\nu) ~~.
$$
And the curvature $R$ is given by:
$$
R=-2(m^2+n^2+mn+m-n){r_0^2\over r^4}A(r)^{-n-1} ~~.          \eqno(12)
$$
We observe that it vanishes like $r^{-4}$ as $r\to\infty$.\noindent
We will study now the geometrical properties of the metric \(6) for
given values of the parameters $m$ and $n$.

To see that the metric \(6) is asymptotically flat it is enough to show
that the metric components behave in an appropriate way at  large
$r$-coordinate values, e.g., $g_{\mu\nu}=\eta_{\mu\nu} + O(1/r)$ as
$r\to\infty$.\noindent
By inspection of the coefficients, we verify that this is so. No matter
which power of $A(r)$, can be written as a binomial series:
$$
A(r)^q=\left(1-2{r_0\over r}\right)^q=1-q{2r_0\over r}+q(q-1)\left(
{2r_0\over r}\right)^2+....                \eqno(13)
$$
{\it Thus, asymptotically flatness is verified for every value of m and n.}

To study the occurrence of {\it true} singularities of the metric \(6),
(not coordinate  system pathologies), it is enough for us to  examine
scalars formed out of the curvature. In particular, the scalar invariant:
$$
I=R_{\alpha\beta\gamma\delta}R^{\alpha\beta\gamma\delta}=
$$
$$
=4r_0^2(r-r_0)^{-2(n+1)}r^{-4+2n}
\cdot \left\{\left({r_0\over r}\right)^2I_1(m,n)+4\left({r_0\over r}
\right)I_2(m,n)+6I_3(m,n)\right\} ~~,            \eqno(14)
$$
where:
$$
I_1(m,n)= 48+56m+41m^2+10m^3+m^4-56n-34mn+
$$
$$
-20m^2n-2m^3n+29n^2+6mn^2+3m^2n^2-8n^3+n^4 ~~,
\eqno(14')
$$
$$
I_2(m,n)= -12-13m-8m^2-m^3+13n+4mn+2m^2n-6n^2+n^3 ~~,   \eqno(14'')
$$
$$
I_3(m,n)= (m+1)^2+(n-1)^2 ~~.                           \eqno(14''')
$$

{}From this expression we observe that the invariant goes
always to zero as $r\to \infty$:

$$
I\longrightarrow O(r^{-6}) ~~,                          \eqno(16)
$$
$$
{\rm as}~~          r\to\infty~~~~~
$$
unless
$m=-1$ and $n=1$. In this case, $I_3(-1, 1)=0$ and also $I_2(-1, 1)=0$.
Thus,
$$
I(-1,1)= 48r_0^4 r^{-8}  ~~.         \eqno(16)
$$
The metric reads particularly simple in this case:
$$
ds^2=-dt^2+dr^2 +r^2A(r)d\Omega^2 ~~.
$$
We recall that for the Schwarzschild metric $(m=n=0)$:
$$
I_{Schw.}={48r_0\over r^6}    ~~.         \eqno(17)
$$
We are also interested in  studying the behavior of the scalar
invariants as $r\to 2r_0$. From expression \(14) we see that
$$
I\longrightarrow O[(r-2r_0)^{-2-2n}]   ~~,               \eqno(18)
$$
$$
{\rm as}~~      r\to 2r_0~~~~~~~~~~~~~~~~~
$$
thus, for not  occurring  a singularity at $r=2r_0$ we must have:
$$
n\leq-1,~~~ no~ singularity~ at~ r=2r_0  ~~.      \eqno(19)
$$
This condition can also be obtained from asking a  non-singular
behavior of the scalar curvature $R$, given by \eq{12}. Notice that
$n\leq-1$ makes $g_{\vartheta\vartheta}$ singular at the horizon.
However, this is only a coordinate singularity since the scalar
invariants, as we have seen, are all finite on the horizon.

One additional non-singular case is given when the term between
curly brackets in \eq{14} vanishes, i. e.,
$$
m=n=0; ~~~ the ~ Schwarzschild~ metric ~~.         \eqno(20)
$$
The other interesting value of the radial coordinate to
study is $r=0$. In this case, we see that:
$$
I\longrightarrow O(r^{-6+2n})  ~~,                      \eqno(21)
$$
$$
{\rm as}~~     r\to 0~~~~~~~~~~
$$
except when $I_1=0$, that is for $n=-m=2$. In this case also
$I_2=0$, then
$$
I(-2, 2)= {48r_0^2\over r^6}  ~~,                      \eqno(22)
$$
$$
{\rm as}~~     r\to 0~~~~~~~~~
$$
which is the same as in the Schwarzschild case. In fact,
it is easy to show that for $n=-m=2$ the the metric \(6)
can be carried into Schwarzschild form. The transformation
$\chi= r-2r_0$ and the identification of $-r_0$ with the
mass $M$ make the job.

We would like now to study the occurrence of an event horizon
at $r=2r_0$.
Let us first observe that the Killing vector $\xi_{(t)}=
\sqrt{-g_{00}}\partial_t=A(r)^{m+1\over2}\partial_t$
becomes null
at $r=2r_0$ when $m+1>0$. We can thus study the outgoing null
geodesics
from $r\geq 2r_0$ and see under which conditions $r=2r_0$
is an outgoing null surface.

The first integral of the geodesics motion (related to the time
and two angular variables) in our spherically symmetric gravitational
field can be written as:
$$
\vartheta={\pi\over 2} ~~,      \eqno(23)
$$
$$
r^2A(r)^n{d\varphi\over d\lambda}= J   ~~,    \eqno(24)
$$
$$
A(r)^{m+1}{dt\over d\lambda}= E  ~~,      \eqno(25)
$$
$$
\left( {dr\over d\lambda}\right)^2= A(r)^{-n+1}
\{ E^2A(r)^{-m-1}-J^2r^{-2}A(r)^{-n}+\epsilon \} ~~,     \eqno(26)
$$
where $\epsilon=0,~\pm 1$ for null, spacelike and timelike
geodesics respectively.

We can describe the radial part of motion in terms of the
effective potential. Then, for null geodesics, we define the
impact parameter $b$ as:
$$
b={J\over E} ~~.      \eqno(27)
$$
The critical impact parameters $b_c$ for which photons with $b>b_c$
can escape to infinity and with $b<b_c$ are absorbed by the black
holes, can be found to be :
$$
{\partial V_{eff}\over \partial r}\biggr\vert _{r_c}=0~~,
{}~~~~~{dr\over d\lambda}(b_c)\biggr\vert _{r_c}=0~~,
{}~~~~~{\partial^2 V_{eff}\over \partial r^2}\biggr
\vert _{r_c}<0 ~~.        \eqno(28)
$$
Thus, the radial coordinate of the critical periastrom $r_c$ is:
$$
{r_c\over r_0}= 3+m-m ~~,     \eqno(29)
$$
and
$$
b_c=r_0 {(3+m-n)^{m-n+3\over 2} \over (1+m-n)^{m-n+1\over 2}} ~~.\eqno(30)
$$
An observer at rest in our gravitational field measures the velocity of a
photon relative to his orthonormal frame \refto{8}:
$$
v_{\hat \phi}={\sqrt{g_{\phi\phi}}d\phi/d\lambda\over
\sqrt{-g_{00}}dt/d\lambda}={b\over r}A(r)^{m-n+1\over 2}  ~~.    \eqno(31)
$$
A photon at $r<r_c$ will eventually scape to infinity instead of being
trapped by the black hole at $r=2r_0$ if $v_{\hat r}$ is positive and:
$$
\sin\delta < {b_c\over r} A(r)^{m-n+1\over 2}  ~~,
$$
where $\delta$ is the angle between the propagation direction and radial
direction.

Thus, we conclude that the surface $r=2r_0$ will act as an event horizon
whenever:
$$
m-n+1>0   ~~.  \eqno(32)
$$

An alternative derivation of the horizon properties of the surface $r=2r_0$
can be obtained by the study of the outgoing radial  null geodesics.
In fact, the time spent by a photon emitted at $r_i$ to reach $r_f$ as
measured by an observer at infinity is given by
$$
\Delta t= \int_{t_i}^{t_f}{dt}=\int_{r_i}^{r_f}{A(r)^{n-m+2\over 2}dr}=
\left[rA(r)^{n-m\over 2}\right]_{r_i}^{r_f}+O\left(A(r)^{{n-m\over 2}+1}
\right).
$$
when $m-n\geq 0$ as $r_i \to 2r_0$ the photon will need a $\Delta t \to
\infty$ to leave the horizon neighborhood, thus indicating the presence
of an event horizon at $r=2r_0$.
For $n\leq -1$ (eq\(19)) we have not singularities on the surface $r=2r_0$.
However, $g_00$ diverges there thus giving an infinite horizon area.
This is only a purely geometrical divergence bringing no physical
consequences.
In fact, we have seen that the surface $r=2r_0$ effectively acts as an
event horizon with respect to null rays.
For massive particles there neither any inconvenient to enter in to the
black hole in a finite proper time since its effective potential,
$V_{eff}$:
$$
E^2-V_{eff}=\left({dr\over d\tau}\right)^2= A(r)^{1-n}\left[E^2A(r)^{-(m+1)}
-{J^2\over r^2}A(r)^{-n}-1\right],
$$
remains bounded at and outside the horizon. Besides, tidal effects on the
horizon are finite since curvature tensors are well behaved there.

\head{3. Astrophysical applications and discussion}

For not only dealing with the mathematical aspects of the solution and to
obtain further restrictions on $m$ and $n$,
we will briefly study  some astrophysical consequences. In particular, we
will compute some physical quantities and  show how much different they
are from the General Relativistic results. Thus, confirming that
metric\(1) is indeed not the Schwarzschild one.

 Nature has the
final word to decide between mathematical models. Thus, to see if
the family of solutions of Brans-Dicke equations could represent
nature, we can start by computing its post Newtonian parameters (PPN). For
an static spherically symmetric metric we can write the PPN metric as
\refto{8}
$$
ds^2=-\left[1-2\left({M\over r}\right)
+2\beta\left(M\over r\right)^2\right]
dt^2+\left[1+2\gamma\left({M\over r}\right)\right]
(dr^2+r^2d\Omega^2) ~~,  \eqno(33)
$$
where $\beta$ and $\gamma$ are two of the ten PPN parameters measuring,
respectively, the amount of nonlinearity in the superposition law for
$g_{00}$ and the amount of space curvature produced by the unit rest mass.

By transforming our metric \(6) to isotropic radial coordinates, $\bar r$,
$$
r=\bar r(1+\bar r_0/ \bar r)^2 ~~; ~~~~~ \bar r_0=r_0/2 ~~,   \eqno(34)
$$
we find \refto{1}
$$
d\bar s^2=-\left({1-\bar r_0/ \bar r  \over 1+\bar r_0/ \bar r}\right )^
{2(m+1)} dt^2 + (1+\bar r_0/ \bar r)^4
\left({1-\bar r_0/ \bar r  \over 1+\bar r_0/ \bar r}\right )^{2n}(d\bar
r^2+\bar r^2d\Omega^2) ~~.    \eqno(35)
$$
By expanding the coefficients of this metric and comparing them to those of
\eq {33} we obtain:
$$
\beta=1 ~~;~~~~~ \gamma={1-n\over m+1} ~~; ~~~~~ M=(m+1)r_0 ~~.  \eqno(36)
$$
Thus, when $m\to -n$ we have agreement with the solar system experiments.
In particular, results of time delay measurements gives \refto{5}:~~
$|\gamma -1|<10^{-3}$.

To find observational differences between metric \(6) with $m\to -n$
(see \(58)), and the Schwarzschild one, we must, then, look at strong
gravitational field effects. We study some of  such effects in accretion
disks, scattering of photons  and Hawking radiation.

The standard model of galactic hard X-ray sources is a binary stellar system
formed by a normal star transferring  matter onto its companion star, which
is a compact object. This matter, falling inward in quasi-circular orbits,
will form an accretion disk, which will emit the observed X-rays.


The friction due to viscosity will generate heat, which is radiated away
through the disk surfaces. This energy is supplied by the loss of the total
energy of the gas, while going through the disk, down to the last stable
circular orbit. After this, the gas would fall almost without radiating
\refto{Stroeger}.

Using Schwarzschild's metric (see ref.\refto{Bardeen} also
for the Kerr case),
the last stable circular orbit has a radial coordinate $r_c=6M$, where $M$
is the black hole mass. At this $r_c$ the energy "at infinity"
per rest energy
is $E_c=(8/9)^{1/2}$. If we take $E\cong 1$ at the external radius of the
disk and a steady flux of matter (or its temporal average) the total
luminosity of the accretion disk will be:
$$
L=(1-E_c)\dot M  ~~,           \eqno(100)
$$
with $\dot M=$ mass per unit time entering the disk.

When we compute $E_c$ from metric \(6) we obtain (for $-n$ large):
$$
{L^{BD}\over L^{GR}}=0.958 \eqno(74)
$$
Another potentially observable quantities are the orbital frequency
of the last stable circular orbit (as seen by an observer at infinity)
and the redshift at infinity  which are given by\refto{Lousto}
$$
{\nu^{BD}_c\over \nu^{GR}_c}=0.931~~;~~~~
{Z^{BD}_c\over Z^{GR}_c}=0.936~~, ~~~~n\to -\infty \eqno(75)
$$

As we have seen, the metric \(6) is in agreement with the solar system
observations  and experiments when $m\to -n$. In addition, $n\leq -1$ for
having a regular horizon. Thus, Eqs. \(74)-\(75) give results close to
those of General Relativity.

Unfortunately, present uncertainties in the modeling and observation of
accretion disk do not provide accurate enough data to discriminate between
metric \(58) and the Schwarzschild one.

{}From the study of null geodesics we made in the last section we can obtain
the total scattering cross section  for photons:
$$
\sigma = \pi b_c^2=\pi {M^2\over (m+1)^2}{(3+m-n)^{m-n+3}\over (1+m-n)^
{m-n+1}} ~~.       \eqno(37)
$$
Let us observe that $m=n$ gives the same results as for a Schwarzschild
black hole. This is so, because when $m=n$ the metric \(6) can be written
conformal to the Schwarzschild one, i.e. $ds^2=A(r)^nds_{Schw}^2$ and
light rays do not "feel" conformal factors.

When we compare this cross section to the General Relativistic result,
$\sigma^{GR}= 27\pi M^2$, in the case $m\to -n$ and $n\leq -1$, we find
that, again, the results are close to those produced by the
Schwarzschild metric.
As $n$ goes to more negative values we have a quick convergence to the
asymptotic value:
$$
{\sigma^{BD}\over\sigma^{GR}}=1.095 ~~,~~~~n\to -\infty       \eqno(78)
$$

{}From the results above one sees that the Brans-Dicke gravitational field
studied seems to be weaker than the Schwarzschild one. This conclusion
will be reinforced when we   compute the surface gravity on the horizon.
Here we find the relatively strongest difference from the Schwarzschild's
results. The surface gravity plays an important role when one studies
the thermodynamics of black holes because it is related to the temperature
associated to  quantum effects close to the horizon. For a static
spherically symmetric system it is given by:
$$
K=-{1\over 2} {g'_{00}\over \sqrt{-g_{00}g_{rr}}}=(m+1){r_0\over r^2}
A(r)^{m-n\over 2}  ~~.    \eqno(50)
$$
When we evaluate it at $r=2r_0$ we find:
$$
K_H = \cases{0,  &for $m>n$ \cr
\infty, &for $m<n$\cr
{m+1\over4r_0}={1\over 4M}=k_{Schw}, &for $m=n$ \cr} ~~, \eqno(51)
$$
thus, we obtain the Schwarzschild value for the conformal case $m=n$.

We can write also our metric  in terms of Kruskal-like variables. To this
 end, let us define first the null variables $\bar u$ and $\bar v$  by
$$
d\bar u=dt-dr^*~~;~~~~~d\bar v= dt+dr^*  ~~,       \eqno(52)
$$
where:
$$
dr^*=rA(r)^{{n-m\over 2}-1}dr  ~~,     \eqno(53)
$$
and then to
$$
u=-\exp(-K_H\bar u)~~;~~~~~ v=\exp(K_H\bar v)  ~~,      \eqno(54)
$$
where $K_H$ is the surface gravity evaluated at $r=2r_0$.

Finally, we obtain \refto{7}
$$
ds^2=-A(r)^{m+1} K_H^{-2} \exp(-2K_H r^*)dudv + r^2d\Omega^2  ~~.
\eqno(55)
$$
When $r\to 2r_0$ and $m=n$  we have
$$
g_{uv}(2r_0)=e^{m+1}\left(4r_0\over m+1\right)^2    ~~.   \eqno(56)
$$
So, metric coefficient are finite on the horizon.

\bigskip

When we bring  together all the conditions for having a regular black
hole, we obtain: $n\leq -1$ for the horizon not being a singular surface
(\eq{19})
and $m-n+1>0$ for $r=2r_0$ acting as an event horizon (\eq{32}). If in
addition, we ask that the solution should be in agreement with the
observations carried out in the solar system, the PPN parameters should
coincide with those of General Relativity with great precision. As we have
seen, this is achieved when $|\omega|\to \infty$, i. e. $m+n\cong 0$.
This, in turn, gives a constant  scalar field  outside the horizon
(see \eq{7}). It is notably that in this case, the Ricci tensor has one
of its components different from zero, i. e.
$$
R_{11}=2n(2-n){r_0^2\over r^4}\left(1-2{r_0\over r}\right)^{-2}
 ~~.   \eqno(57)
$$
This is so, because in spite of  the scalar field $\phi$ being constant
and thus its derivatives going to zero, the coupling constant $|\omega|$
goes to infinity in such a way that the product appearing on the right
hand side of the field equations
\(4) gives a finite value, i. e. (Eq.\(57)).
This fact has very important consequences for the Hawking theorem \refto{3}
establishing the identity of Brans-Dicke and General Relativity Black Holes.
Indeed, the limiting case $m+n\to 0^-$  is contained within our family of
solutions and it is well defined (For example $m=-2$, $n=2$ gives
the Schwarzschild metric). In this  black hole solution \(58):
$$
ds^2=-A(r)^{1-n}dt^2+A(r)^{n-1}dr^2+r^2A(r)^nd\Omega^2 ~~,
$$
the parameter $n$ plays the role of a classical Brans-Dicke hair. It
has its origin in the particular coupling of the Brans-Dicke scalar field.
Their effects at large distances can be absorbed in a redefinition of the
mass of the black hole and thus as we have seen, at Post-Newtonian  level
this metric coincides with  the Schwarzschild one. However, as we study
strong gravitational field effects, their results are  dependent on the
value of $n$. In some sense, thus, $n$ has an intermediate range of action.

When $\phi$ is not constant, black hole solutions are still possible due to
the fact that the surface integral
$\int{(\varphi^2)^{,\alpha}d\Sigma_{\alpha}}$ (where $\varphi=\phi-\phi_0$)
assumed to vanish\refto{3,5} (under the implicit supposition of
$T_{\mu\nu}l^\mu l^\nu\geq0$), here gives a non - zero contribution,
i. e., $4\pi r_0\phi_0^2(m+n)<0$, thus compensating the positive value
of the integral $\int{(\varphi_{,\alpha})^2\sqrt{-g}dx^4 }$.
This is indeed so due to the particular form of the scalar field  \eq{7},
 which produces a stress tensor that violates the weak energy condition
(with $(m+n)<0$ ensuring regularity of the field on the horizon).
The finiteness of the surface integral can be understood
 by the fact that $g_{\vartheta\vartheta}$ for $n\leq -1$
 diverges on the horizon, thus
producing a finite result when multiplied by the vanishing scalar field
terms and integrated over the horizon surface.

It is worth to stress that as $m+n\to 0^-$ and $n\leq-1$, then
$\omega\to-\infty$.
This is perfectly acceptable because there is no theoretical reason to
restrict $\omega$ to positive values\refto{ompos} and experiments are
consistent with $|\omega|\gsim500$.
For the allowed range of values of the parameters $m$ and $n$ (given by eqs.
\(19) and \(32)), $-\infty<\omega<-4/3$.
Let us remember that the string theory
selects the value\refto{2} $\omega_S=-1$; while the graceful exit problem
is solved for \refto{1} $\omega_{EI}\leq 20$.

The no hair theorem can be overcome because the weak energy condition is
violated by the energy momentum tensor of the Brans-Dicke field. In fact,
$$
T_{00}=n(2-n){r_0^2\over r^4}(1-2r_0/r)^{m-n}~~,     \eqno(59)
$$
that for $n\leq -1$, takes always negative values (independent of the limit
$m\to -n$).

Another interesting result is that for the subfamily  \(58), in particular
(in general see  \eq{32}),  the surface
gravity will be zero (see \eq{51}). Hence, these black holes are truly
"black", even at the semiclassical level, in the sense that not Hawking
radiation is  expected to take place here.
It is worth to remark here that the divergence of the horizon surface
not only does not affect the computation of relevant physical quantities,
but can also be interpreted,  with regards to the thermodynamics of black
holes, as suggesting an infinite entropy for our black hole solutions.
This, in turn, is consistent with its associated semiclassical zero
temperature.

We remark that this kind of analysis can be also carried out for the
generalization of the  Kerr-Newmann metric in the Brans-Dicke theory
\refto{6}.

The problem of stability of  solution \(58) is now under study by the
present authors, but we can advance some comments: Matsuda \refto{Ma} has
found, studying the spherically gravitational collapse of a
star in Brans-Dicke theory, that it does not necessarily produce a
Schwarzschild black hole, but can also produce the black hole solution
given by metric \eq{58}.

The radiation of the scalar field will be damped by a factor \refto{3, 5}
$(2+\omega)^{-1}$, which vanishes as $|\omega|$ goes to infinity.
Thus, we think that metric \(58) is a viable candidate to represents the
black holes in nature.

\bigskip
\bigskip
\vskip 12pt
\noindent
{\it Acknowledgements}

\noindent
The authors are very grateful to G. Arcidiacono for calling our attention on
solutions Eq.\(6) and J. Audretsch for wise advise.
We are indebted to A. Economou for useful conversations
and computational assistance. \mangos

\vfill

\references

\refis{1} D. La and P. J. Steinhardt, 1989, {\it Phys. Rev. Lett.}
          {\bf 62}, 376; E. W. Kolb, 1990 (Sweden, June 1-8),
	  {\it "First Order Inflation"}, Proceeding of the Nobel Symposium
	  \# 79; M. S. Turner, 1991 (September 1-8), {\it "First Order
	  Inflation"}, Lectures at the Erice School "Daniel Chalonge".\par

\refis{2} C. Lovelace, 1984, {\it Phys. Lett.} {\bf B 135}, 75; E. S.
          Fradklin and A. A. Tseytlin, 1985, {\it Phys. Lett.}
	  {\bf B 158}, 316; C.G. Callan, D. Friedan, E. J. Martinec
	  and M. J. Perry, 1985, {\it Nucl. Phys.} {\bf B 262}, 593;
	  A. Sen, 1985, {\it Phys. Rev.} {\bf D 32}, 2102;
          A. Sen, 1985, {\it Phys. Rev. Lett.} {\bf 55}, 1846;
          C. G. Callan, I. R. Klebanov and M. J. Perry, 1986,
	  {\it Nucl. Phys.} {\bf B 278}, 78.\par

\refis{3} K. S. Hawking, 1972, {\it Commun. Math. Phys.} {\bf 25},
          167.\par

\refis{4} P. Jordan, 1959, {\it Phys}, {\bf 157}, 128; C. H. Brans and
          R. H. Dicke, 1961, {\it Phys.Rev.} {\bf 124}, 925.\par

\refis{5}  C. M. Will, 1981, {\it ``Theory and Experiment in
           Gravitational Physics"}, Cambridge Univ. Press;
	   C. M. Will, 1984, {\it Phys. Rep.} {\bf 113}, 345.\par

\refis{6} K. D. Krori and D. R. Bhattacharjee, 1982, {\it J. Math. Phys.}
          {\bf 23}, 637.\par

\refis{7} C. O. Lousto and N. S\'anchez, 1989, {\it Phys. Lett.} {\bf
          B 220}, 55.\par

\refis{8} C. W. Misner, K. S. Thorne and J. A. Wheeler, 1973, {\it
          Gravitation}, Freeman \& Co, San Francisco.\par

\refis{Bardeen} J. M. Bardeen Press and S. A. Teukolsky, 1972, {\it Ap. J.}
                {\bf 178}, 347.\par

\refis{Stroeger} W. R. Stroeger, 1980, {\it Ap. J.} {\bf 235}, 216.\par

\refis{Lousto} C. O. Lousto, 1986, {\it Rev. Mex. Astr. Astrof.},  Vol.{\bf
               13}, 3; C. O. Lousto and H. Vuchetich, 1988, {Proceedings of
	       the SILARG VI}, M. Novello Ed., World Sci., 260.\par

\refis{LW} K. Lee and E. J. Weinberg, 1991, {\it Phys. Rev. }  {\bf D 44},
           3159.\par

\refis{GHS} D. Garfinkle, G. T. Horowitz, and A. Strominger, 1991, {\it Phys.
            Rev.} {\bf D 43}, 3140.\par

\refis{B} J. D. Bekenstein, 1975, {\it Ann. Phys. (N. Y.)} {\bf 91}, 75.\par

\refis{CDKO} B. A. Campbell, M. J. Duncan, N. Kaloper, and K. A. Olive, 1990,
             {\it Phys. Lett.} {\bf B 251}, 34.\par

\refis{BK} K. A. Bronnikov and Yu. N. Kiraev, 1978, {\it Phys. Lett.}
{\bf 67A}, 95.\par

\refis{Ma} T. Matsuda, 1972, {\it Prog. Theor. Phys.} {\bf 47}, 738.\par

\refis{ompos} L .L. Smalley and P. B. Eby, 1976, Nouvo Cim., {\bf B35}, 54.
              N.T.Bishop, 1976, MNRAS, {\bf 176}, 241 .

\endreferences

\end